\documentclass[prd,twocolumn,showpacs,amsmath,amssymb,superscriptaddress,preprintnumbers,nofootinbib,showkeys]{revtex4-2}
\usepackage{graphicx}

\begin{document}

\newcommand{\nablab}{{\mathop {\rule{0pt}{0pt}{\nabla}}\limits^{\bot}}\rule{0pt}{0pt}}

\title{Nonlocal extension of causal thermodynamics of the isotropic cosmic fluid}

\author{Alexander B. Balakin}
\email{Alexander.Balakin@kpfu.ru} \affiliation{Department of
General Relativity and Gravitation, Institute of Physics, Kazan
Federal University, Kremlevskaya str. 16a, Kazan 420008, Russia}

\author{Alexei S. Ilin}
\email{alexeyilinjukeu@gmail.com} \affiliation{Department of
General Relativity and Gravitation, Institute of Physics, Kazan
Federal University, Kremlevskaya str. 16a, Kazan 420008, Russia}

\date{\today}

\begin{abstract}
We establish the nonlocal generalization of the Israel-Stewart model for the relativistic causal thermodynamics of the cosmic fluid, which evolves in the homogeneous isotropic Universe. Based on the second law of thermodynamics we derive the integro-differential master equation for the nonequilibrium pressure scalar and reduce it to the differential equation of the second order in time derivatives. We show that this master equation can be considered as the relativistic analog of the Burgers equation, which is known in the classical theory of viscoelasticity. We obtain the nonlinear key equation, which contains the energy density scalar only, and analyze two exact solutions of the model. The effective temperature is considered to be associated with the barotropic equation of state of the cosmic fluid.

\end{abstract}
\maketitle

\section{Introduction}

\subsection{Prologue}

The relativistic causal thermodynamics elaborated by Israel and Stewart \cite{IsraelStewart} presents the remarkable page of the story of the development of the irreversible thermodynamics. The second law of the phenomenological thermodynamics, which declares that the entropy production $\sigma$ of a closed physical system should be non-negative $\sigma \geq 0$, is the basic element of all versions of the relativistic phenomenological thermodynamics. The entropy production scalar $\sigma$ is defined as the covariant divergence of the entropy flux four-vector $S^k$, i.e., $\sigma=\nabla_kS^k$.
 The difference between the versions of the irreversible thermodynamics, which have been formulated till now, is encoded in the structure of the four-vector $S^k$. The Eckart's version of this quantity \cite{Eckart} is known to have the form
\begin{equation}
S^k_{(\rm Eckart)} = s_0 n U^k + \frac{1}{T} q^k \,,
\label{1}
\end{equation}
where $n$ is the scalar of particle number density, $T$ is the temperature, $s_0$ is the scalar of entropy per one particle, $U^k$ is the timelike medium velocity four-vector, and $q^k$ is the spacelike heat-flux four-vector. The Israel-Stewart version of the entropy flux four-vector is
$$
S^k_{(IS)} = S^k_{(\rm Eckart)} +  \frac{1}{T} q^l \left[\delta^k_l \alpha_0 \Pi  + \alpha_1 \Pi^k_{\ l(0)} \right] -
$$
\begin{equation}
- \frac{1}{2T} U^k \left[\beta_0 \Pi^2 - \beta_1 q^m q_m + \beta_2 \Pi^{mn}_{(0)} \Pi_{mn(0)}\right] \,.
\label{12}
\end{equation}
In fact, Israel and Stewart have added all the possible terms of the second order with respect to the non-equilibrium quantities: the heat-flux four-vector $q^k$, the scalar part $\Pi$ of the non-equilibrium pressure $\Pi_{ik}$, and its traceless shear part $\Pi_{mn(0)}$. We use standardly the decomposition $\Pi_{ik}= \Pi_{ik(0)} + \frac13 \Delta_{ik} \Pi$, where $\Delta_{ik} \equiv g_{ik}-U_iU_k$ is the projector. New phenomenological parameters $\alpha_0$, $\alpha_1$, $\beta_0$, $\beta_1$ and $\beta_2$ were considered as functions of the temperature $T$.

The standard Gibbs equation includes the scalar $s_0$
\begin{equation}
De + P D\left(\frac{1}{n} \right) = T Ds_0 \,,
\label{2}
\end{equation}
where $e$ describes the energy density per one particle $e= \frac{W}{n}$; $W$ is the energy density scalar; $P$ is the isotropic equilibrium Pascal pressure; $D$ is the convective derivative defined as $D=U^k \nabla_k$ with the covariant derivative $\nabla_k$. The listed quantities form the stress-energy tensor of the medium
\begin{equation}\label{3}
T^{ik} = W U^i U^k + U^i q^k + U^k q^i - \Delta^{ik} P + \Pi^{ik} \,.
\end{equation}
The heat flux four-vector is orthogonal to the velocity four-vector, $q_k U^k{=}0$, and the symmetric non-equilibrium pressure tensor satisfies the condition $\Pi^{ik}U_k{=}0$.

\subsection{Relativistic causal thermodynamics of the isotropic homogeneous Universe}

Numerous applications of the causal thermodynamics to the cosmological models of the Friedmann type (see, e.g., \cite{F1} - \cite{F13}) were based on the evident fact that due to the claimed spacetime symmetry one has to put the spacelike heat-flux four-vector to zero, $q^k \equiv 0$, and to consider the tensor $\Pi_{mn(0)}$ to be vanishing. Also, all the thermodynamic quantities are considered to be the functions of the cosmological time only, and the velocity four-vector is chosen to be of the form $U^i = \delta ^i_0$. These requirements are connected with the idea of inheritance of the symmetry of the spacetime metric
\begin{equation}\label{F}
ds^2 = dt^2 - a^2(t)\left[dx^2+ dy^2+ dz^2 \right] \,, \quad c=1 \,,
\end{equation}
by all the physical quantities: the energy density scalar, Pascal pressure, etc.
With these assumptions, the entropy production scalar is calculated to have the form
\begin{equation}
\sigma = \frac{\Pi}{T}  \left[\frac13 \Theta - \beta_0 D \Pi - \frac12 T \Pi \nabla_l \left(\frac{\beta_0 U^l}{T} \right) \right] \geq 0 \,,
\label{019}
\end{equation}
and the non-equilibrium pressure $\Pi$ appeared as the solution to the equation
\begin{equation}
\beta_0 D \Pi {+} \Pi \left[\frac{1}{9\zeta} {+} \frac{T}{2} (\Theta {+}D) \left(\frac{\beta_0}{T} \right)\right] = \frac13 \Theta  \,.
\label{14}
\end{equation}
The phenomenologically introduced function $\zeta$ plays the role of the bulk viscosity coefficient; the scalar $\Theta \equiv \nabla_kU^k$ is the expansion scalar of the medium flow.

\subsection{Motivation and structure of the work}

\subsubsection{On the classical Newton and Maxwell viscoelastic models}

According to the standard terminology accepted in the classical theory of viscoelasticity (see, e.g., \cite{visco}), and in the theory of irreversible thermodynamics of fluid (see, e.g., \cite{JCL}), when $\beta_0{=}0$, the equation (\ref{14}) describes the analog of the model of the Newton fluid, for which the stress tensor  is proportional to the time derivative of the strain tensor. In the classical linear theory of viscoelasticity the strain tensor $\varepsilon_{\mu \nu} = \frac12 \left(\frac{\partial {\cal U}_{\nu}}{\partial x^{\mu}} + \frac{\partial {\cal U}_{\mu}}{\partial x^{\nu}} \right)$ contains the displacement three-vector ${\cal U}^{\nu}$ (Greek indices take values $1,2,3$). For illustration only, we use  for this object the term $\epsilon$ without indices. Since in the classical viscoelasticity the three velocity of the medium $V^{\nu}$ is equal to the time derivative of the displacement vector, i.e., $V^{\nu} = \dot{{\cal U}}^{\nu}$, we see that $\dot{\varepsilon}_{\mu \nu} = \frac12 \left(\frac{\partial V_{\nu}}{\partial x^{\mu}} + \frac{\partial V_{\mu}}{\partial x^{\nu}} \right)$, thus, for the Newton fluid we can use the compact formula  $\pi = \eta \dot{\epsilon}$, where symbol $\pi$ indicates the stress tensor, and the $\eta$ is the viscosity coefficient.
When $\beta_0 \neq 0$, the equation (\ref{14}) describes the analog of the Maxwell model of viscoelasticity, for which $\tau \dot{\pi} + \pi = \nu \dot{\epsilon}$. In other words, inheriting the properties of the classical Newton and Maxwell fluids the Israel-Stewart model has been constructed based on the analog of the so-called tensor of deformation rates only, $\dot{\varepsilon}_{\mu \nu}$.

\subsubsection{Classical fluids with elastic properties}

The analog of the Hooke's term, which in classical viscoelasticity is of the form $\pi = E \epsilon$, does not appear in the equation (\ref{14}). However, it would be interesting to study the analog of the classical model, associated with the constitutive equation of the type $\tau \dot{\pi} {+} \mu \pi {=} \nu \dot{\epsilon} {+} E \epsilon$, which contains both quantities $\dot{\epsilon}$ and $\epsilon$. This constitutive equation is extensively used in the transient and steady-state rheologies, as well as, in the theory of viscoelastic media and materials (see, e.g., \cite{Earth}). There is the direct link between this equation and the equation obtained in the Burgers model of viscoelasticity (see, e.g., \cite{Burg1,Burg3})
\begin{equation}
\tau \ddot{\pi} {+} \mu \dot{\pi} {+} \omega \pi = \nu \ddot{\epsilon} {+} E \dot{\epsilon} \,.
\label{Burgers1}
\end{equation}
The equation (\ref{Burgers1}) is of the second order in time derivatives for the stress $\pi$ and contains the strain $\epsilon$ with the first and second time derivatives; the strain itself is absent.

\subsubsection{Towards the covariant description of the fluid with viscoelastic properties}

 When one establishes the relativistic  theory of elasticity and viscoelasticity, the problem appears how to define the covariant version of the displacement four-vector, or of the covariant strain tensor (see, e.g., \cite{relVISC1,relVISC2,relVISC3}), however, there are no problems with the formulation of the covariant analog of the tensor $\dot{\varepsilon}_{\mu \nu}$. In this sense, when we address to the covariant generalization of the Burgers equation, we avoid the mentioned problem and open a new window in the description of relativistic fluids with viscoelastic rheologic properties.
 Why it could be interesting for cosmology? There are two motives for this interest, and the first one is connected with the search for adequate equation of state of the cosmic fluid.

 Modern theories of the Universe evolution operate with the so-called dark fluid; for the description of the late-time Universe evolution cosmologists, based on recent observations, consider the dark fluid to be a two-component substratum, which contains the dark energy with negative pressure and the pressureless dark matter. When scientists try to reconstruct the evolution of the cosmic fluid in early Universe, e.g., during the inflation, reheating, etc., they use the well-known instrument: modeling of the equations of state, which link the fluid pressure and energy density. These equations of state can be divided into three classes. The first class contains the functional equations, for instance, the barotropic ones with the time dependent coefficients (see, e.g., \cite{B1,B2}). The second class includes the differential constitutive equations (see, e.g., \cite{D1,D2,D3,D4}). In the third class the integral representation of the equation of state is used (see, e.g., \cite{BI}); the models of this class can be indicated as the nonlocal ones. In the process of modeling of the early Universe evolution we can imagine that, being rather dense, the cosmic fluid was a rheologic substratum and had viscoelastic properties. In this sense the nonlocal equations of state of the cosmic fluid attract attention.
In this work we established the nonlocal extension of the Israel-Stewart theory, thus formulating the constitutive equation of the cosmic fluid in the integral representation. Surprisingly, this constitutive equation for the non-equilibrium pressure happened to be equivalent to the covariant analog of the Burgers equation.
In other words, the established nonlocal model gives the possibility to describe one of the variants of the isotropic relativistic viscoelastic cosmic fluid of the Burgers type.

The second motif of the interest to the formulated problem is the following. In the proposed nonlocal approach the entropy production scalar $\sigma$, which is the important function characterizing the thermodynamic state of the fluid, contains  the integral over the non-equilibrium pressure $\Pi$. This means that the value of the entropy production at the moment $t$ depends not only on the instantaneous values of the state functions, but also it depends on all the prehistory of the evolution of the function $\Pi$, which can be positive, negative and equal to zero during some time intervals. We deal, in fact, with some cumulative thermodynamic effect, which can be related to the simplest variant of the fluid memory.

Mention should be made that in our work we consider the one-component cosmic fluid. In reality the cosmic fluid is multi-component, however, for basic stages of the Universe evolution one can select one dominating component and consider the corresponding truncated model. Of course, one-component representation of the cosmic fluid is the first step in the development of this theory, and we hope to extend it in future.
Depending on the choice of the constitutive parameters, this cosmic fluid may be indicated as the dark energy, phantom-like, ultrarelativistic, perfect fluid, etc., but in the sake of generality we use the unified term cosmic fluid in the text of the paper.

The paper is organized as follows. In Section II we describe the formalism of the nonlocal extension of the Israel-Stewart theory based on the Friedmann spacetime platform. In Section III we derive the key equations of the model and analyze them using two examples of exact solutions. Section IV contains conclusions.

\section{Nonlocal extension of the Israel-Stewart theory}

\subsection{The formalism}

\subsubsection{Evolutionary equation for the non-equilibrium pressure $\Pi$ }

We consider now the decomposition of the entropy flux four-vector for the spatially homogeneous isotropic cosmological model in the following form:
\begin{equation}
S^k =  s_0 n U^k - \frac{1}{2T} U^k \beta_0 \Pi^2 + \frac12 \gamma_0 nU^k \left(D^{-1} \Pi \right)^2  \,,
\label{17}
\end{equation}
where $\gamma_0$ is a constant. The term $D^{-1} \Pi$ is defined as follows:
\begin{equation}
D \left(D^{-1} \Pi\right) = \Pi \,, \quad D^{-1} \Pi = \int d\tau  \Pi  \,, \quad d\tau = U_k dx^k \,.
\label{18}
\end{equation}
The coefficient $\frac12 \gamma_0 nU^k$ is chosen so that its divergence is equal to zero due to the conservation law for the particle number
\begin{equation}
\nabla_k \left(nU^k \right)=0 \,, \quad  \Rightarrow Dn + n \Theta =0  \,.
\label{T10}
\end{equation}
Now the entropy production scalar takes the form
\begin{equation}
\sigma = \frac{\Pi}{T}  \left[\frac{\Theta}{3} {-} \beta_0 D \Pi {-} \frac{T}{2} \Pi \nabla_l \left(\frac{\beta_0 U^l}{T} \right)  {+}  \gamma_0 n T \left(D^{-1} \Pi \right) \right] \,.
\label{19}
\end{equation}
This quantity is non-negative, when
\begin{equation}
\frac{\Pi}{9\zeta} = \frac{\Theta}{3} {-} \beta_0 D \Pi {-} \frac{T}{2} \Pi \nabla_l \left(\frac{\beta_0 U^l}{T} \right)  {+}  \gamma_0 n T \left(D^{-1} \Pi \right)  \,.
\label{20}
\end{equation}
Clearly, the equation (\ref{20}) recovers the definition of the non-equilibrium pressure appeared in the Israel-Stewart theory, if $\gamma_0=0$.

The integro-differential equation (\ref{20}) can be rewritten in the form of differential equation of the second order
\begin{equation}
D\left[\frac{\Pi}{9n \zeta T} {-} \frac{\Theta}{3n T}  {+} \frac{\beta_0}{n T} D \Pi {+} \frac{\Pi}{2n}  \left(\Theta {+} D \right) \left(\frac{\beta_0} {T} \right)\right]  {=} \gamma_0  \Pi.
\label{21}
\end{equation}
We can rewrite this equation in the form
$$
\frac{\beta_0}{n T} D^2 \Pi   {+}  D\Pi \left[D \left(\frac{\beta_0}{n T}\right) {+} \frac{1}{9 n T \zeta} {+}  \frac{1}{2n} \left(\Theta {+} D \right) \left(\frac{\beta_0}{T} \right) \right] {+}
$$
$$
{+} \Pi \left\{{-}\gamma_0 {+} D\left(\frac{1}{9 n T \zeta} \right) {+} D \left[\frac{1}{2 n} \left(\Theta {+} D \right) \left(\frac{\beta_0} {T} \right) \right] \right\}  =
$$
\begin{equation}
= D\left(\frac{1}{3n T} \Theta \right) \,,
\label{22}
\end{equation}
analogous to the Burgers equation.

\subsubsection{Reduced gravity field equations and evolution of the energy density scalar}

The conservation law $\nabla_k T^{ik}=0$ can be reduced to
\begin{equation}
\dot{W} + \left[W+P- \frac13 \Pi \right]\Theta   = 0 \,,
\label{5}
\end{equation}
where $\Theta$ can be expressed in terms of the Hubble function $H(t)= \frac{\dot{a}}{a}$, i.e., $\Theta = 3H$. Here and below the dot denotes the derivative with respect to cosmological time.
Then we add the Einstein equation,
\begin{equation}
3H^2 = \kappa W \,,
\label{5v}
\end{equation}
which links the Hubble function and the energy density scalar $W$ ($\kappa \equiv 8 \pi G$). For further analysis we add also the formula for the acceleration parameter $-q(t)$
\begin{equation}
- q(t) \equiv \frac{\ddot{a}}{a H^2} = -\frac12 (1+3\gamma) + \frac{\Pi}{2W}  \,.
\end{equation}
Clearly, the geometric parameters $H$ and $q$ become known, when the state functions $W$ and $\Pi$ are found.

\subsubsection{Effective equation of state}

The scalar $P$ describes the equilibrium Pascal pressure. The total pressure ${\cal P}$ contains both equilibrium and non-equilibrium parts, ${\cal P} = P {-} \frac13 \Pi$.
The known phenomenological approach for the formulation of the equation of state is based on the idea that the total pressure is proportional to the energy density scalar ${\cal P} = w(a(t)) W$, where     $w(a(t))$ is some constitutive coefficient depending on time via the scale factor $a$ (see, e.g., \cite{B2} for details). We use the equivalent scheme: we consider separately the constitutive equation  (\ref{22}) for the non-equilibrium pressure $\Pi$ and the barotropic equation of state $P=\gamma W$ with the constant parameter $\gamma$ for the equilibrium pressure. These two approaches can be linked as follows:
\begin{equation}
w(a(t)) = \frac{P-\frac13 \Pi}{W} = \gamma  - \frac{\Pi}{3W}\,.
\label{5h2}
\end{equation}
Again the constitutive function $w$ is known, when the state functions $\Pi$ and $W$ are found.

\subsubsection{Evolution of the temperature $T$}

Since in the isotropic Universe, which we consider in our model, the heat-flux four-vector vanishes, $q^i=0$, and since the spatial gradient of the temperature is equal to zero, we cannot obtain the standard equation for the temperature evolution by the Cattaneo scheme \cite{Cattaneo}. We need now some additional equation for the temperature depending on the cosmological time only. The corresponding procedure was described, e.g., in the works \cite{F4,F7,T1,T2,T3,B2}. Let us recall the main idea of this procedure. When we consider the energy density $W(n,T)$ and the equilibrium Pascal pressure $P(n,T)$ to be functions of the particle number density $n$ and temperature $T$, we have to require that
\begin{equation}
\frac{\partial^2 s_0 }{\partial n \partial T} = \frac{\partial^2 s_0 }{\partial T \partial n} \,.
\label{T1}
\end{equation}
This means that the entropy is considered as the state function with continuous derivatives up to the second order. Using the Gibbs equation (\ref{2}) and the condition (\ref{T1}), we obtain immediately the compatibility equation
\begin{equation}
n \frac{\partial W}{\partial n} + T \frac{\partial P}{\partial T} = W+P \,,
\label{T2}
\end{equation}
which has to link the functions $W(n,T)$ and $P(n,T)$.

When the equilibrium equation of state for the fluid is linear barotropic, $P = \gamma W$, i.e., the state functions are linked directly and thus, the functions $T$ and $n$ do not take part in the formulation of the equation of state, the functions $T$ and $n$ become the latent ones. Evolution of the function $n(t)$ is predicted by the law (\ref{T10}); we need also the equation for the function $T(t)$.
For the linear barotropic equation of state the energy density scalar has to satisfy the linear differential equation in partial derivatives
\begin{equation}
n \frac{\partial W}{\partial n} + \gamma T \frac{\partial W}{\partial T} = (1+ \gamma) W \,.
\label{T3}
\end{equation}
Solving the characteristic equations
\begin{equation}
\frac{dn}{n} = \frac{dT}{\gamma T} = \frac{dW}{(1+\gamma) W} \,,
\label{T4}
\end{equation}
we obtain the energy density scalar in the following form
\begin{equation}
W = n^{1+\gamma} {\cal F}\left(\frac{T}{n^{\gamma}} \right) \,,
\label{T5}
\end{equation}
where ${\cal F}$ is arbitrary function of its argument.
Equivalently, one can express the temperature as follows:
\begin{equation}
T = n^{\gamma} {\cal G}\left(\frac{W^{\frac{\gamma}{1+\gamma}}}{n^{\gamma}} \right) \,,
\label{T6}
\end{equation}
where ${\cal G}$ is arbitrary function of its argument.
In this formula for the particle number scalar $n(t)$ we can use the result of integration of the equation (\ref{T10}), namely,
\begin{equation}
n(t)=n_0 \left(\frac{a(t_0)}{a(t)}\right)^3 \,, \quad n_0 = n(t_0)\,.
\label{n1}
\end{equation}
It is interesting to attract attention to two special cases.

\noindent
1.  When the function ${\cal G}$ in (\ref{T6}) is linear; then the scalar $n$ disappears, and we obtain
\begin{equation}
T = \left(\frac{W}{\omega_0}\right)^{\frac{\gamma}{1+\gamma}} \Rightarrow \  W = \omega_0 T^{\frac{1+\gamma}{\gamma}}  \,.
\label{T68}
\end{equation}
This is the example of an one-parameter equation of state since the function $n(t)$ disappears from this equation. There are two known particular cases.

\noindent
1.1. If we require $\gamma = \frac13$, i.e., in the state of equilibrium the fluid is ultrarelativistic, we deal with the Stefan-Boltzmann law $W=\varkappa T^4$.

\noindent
1.2. If $\gamma = - \frac13$, i.e., in the state of equilibrium the fluid is phantom-like, we obtain the constitutive law $W=\omega_0 T^{-2}$.

\noindent
2.  When $\gamma {=}{-}1$, we deal with the formulas $W {=} {\cal F}\left(Tn \right)$ with arbitrary function ${\cal F}$.
Now we obtain $W{+}P{=}0$, thus the fluid in the equilibrium state, when $\Pi{=}0$ can be associated with the dark energy. In the non-equilibrium state $W{+} {\cal P} = {-} \frac13 \Pi \neq 0$, and the type of fluid is predetermined by the sign of the non-equilibrium pressure $\Pi$.

Finally, one can say that we obtain the set of four equations (\ref{22}), (\ref{5}), (\ref{5v}), (\ref{T6}), which link four unknown functions $\Pi(t)$, $T(t)$, $W(t)$  and $H(t)$. In addition, we know that the function $n$ is given by $n(t)=n(t_0) \frac{a^3(t_0)}{a^3(t)}$, and $H(t)=\frac{\dot{a}}{a}$, thus, the problem is well posed.

\subsection{Truncated model and the key equation}

Our goal is to derive and analyze the so-called key equation, which includes only one unknown function, say, the energy density scalar $W$. In order to reach some analytic progress in calculations, we can use the ansatz about the structure of the phenomenologically introduced coefficients $\zeta$ and $\beta_0$. First of all, we assume that the bulk viscosity coefficient depends on the temperature $T$ and on the expansion scalar $\Theta$ as follows: $\zeta(T,\Theta) = \frac{\zeta_0}{T \Theta}$, where $\zeta_0$ is some constant. This assumption is motivated by the results of numerous investigations of the bulk viscosity coefficient for various media and materials: generally, this coefficient decreases when the temperature grows. Also we assume that $\beta_0(T) = h_0 \cdot T$, where $h_0$ is some constant.  The second assumption relates to the findings that the relaxation time associated with the coefficient $\beta_0$ grows when the temperature increases. Of course, these assumptions should be considered as an ansatz, which has to be verified later. Now we use the relationship $\Theta = 3H$ and simplify the equation (\ref{22}) distinguishing two cases: $h_0 \neq 0$ and $h_0 = 0$.

\subsubsection{The key equation for the model with $h_0 \neq 0$}

When $h_0 \neq 0$ we obtain from (\ref{22})
$$
\ddot{\Pi} {+} H \dot{\Pi} \left(\frac{1}{3 \zeta_0 h_0 } {+} \frac{9}{2} \right) {+}
$$
$$
+
\Pi \left[\left(\dot{H}{+}3H^2 \right) \left(\frac{1}{3 \zeta_0 h_0 } {+} \frac{3}{2} \right)  {-} \frac{\gamma_0}{h_0 } n \right] =
$$
\begin{equation}
= \frac{H}{h_0 T} \left(\frac{\dot{H}}{H} {+} 3H {-} \frac{\dot{T}}{T}  \right) \,.
\label{cc1}
\end{equation}
In order to simplify this equation,
we introduce the new dimensionless variable $x=\frac{a(t)}{a(t_0)}$ and replace the derivative with respect to time using the relationship  $\frac{d}{dt}= xH \frac{d}{dx}$. We obtain from (\ref{cc1}) the equation
$$
x^2 \Pi^{\prime \prime}(x) + x \Pi^{\prime}(x) \left[\frac{11}{2} + \frac{1}{3\zeta_0 h_0} + x \frac{H^{\prime}}{H} \right] +
$$
$$
+\Pi \left[\left(x \frac{H^{\prime}}{H}{+}3 \right) \left(\frac{1}{3 \zeta_0 h_0 } {+} \frac{3}{2} \right)  {-} \frac{\gamma_0 n_0}{h_0 x^3 H^2 } \right] =
$$
\begin{equation}
= \frac{1}{h_0 T} \left(x\frac{H^{\prime}}{H} {+} 3 {-} x\frac{T^{\prime}}{T}   \right) \,.
\label{x12}
\end{equation}
Here and below the prime denotes the derivative with respect to the variable $x$. Using the Einstein equation (\ref{5v}) we make the replacements $H^2 \to \frac{\kappa}{3} W$, and $\frac{H^{\prime}}{H} = \frac{W^{\prime}}{2W}$. Using the conservation law (\ref{5}) we can write now
\begin{equation}\label{ME2}
\Pi = x W^{\prime}(x) + 3(1+\gamma) W \,.
\end{equation}
Finally, we express the temperature and its derivative via the $W$ and $n$, using (\ref{T6}) and keeping in mind that $n=\frac{n_0}{x^3}$ from (\ref{n1}).
Thus, we obtain the nonlinear equation of the third order for the function $W(x)$ only; this is the key equation, it has the form
$$
x^3 W^{\prime \prime \prime} + x^2 W^{\prime \prime}\left[3\gamma + \frac{21}{2} + \frac{1}{3\zeta_0 h_0} + x \frac{W^{\prime}}{2W} \right]+
$$
$$
{+} xW^{\prime}\left[x \frac{W^{\prime}}{2W} \left(\frac{1}{3 \zeta_0 h_0 } {+} \frac{11}{2} {+} 3\gamma \right) {+} \frac{17{+}9\gamma}{6 \zeta_0 h_0} {+} \frac54 (23{+}15\gamma) \right] {+}
$$
$$
+3(1+\gamma) W  \left(\frac{1}{\zeta_0 h_0 } {+} \frac{9}{2} \right)  =
$$
\begin{equation}
= \frac{1}{h_0 T} \left(x \frac{W^{\prime}}{2W} {+} 3 {-} x\frac{T^{\prime}}{T}   \right) {+}  \frac{3\gamma_0 n_0}{h_0 \kappa x^3}  \left[x\frac{W^{\prime}}{W} {+} 3(1{+}\gamma) \right].
\label{key1}
\end{equation}

\subsubsection{The key equation for the model with $h_0 = 0$}

When $h_0 = 0$ and thus $\beta_0=0$, the equation for $\Pi$ obtained in the framework of the Israel-Stewart theory, is of the type typical for the Newtonian fluid, $\Pi = \frac13 \Theta$. However, the nonlocal extension of the Israel-Stewart theory, presented in our paper, gives the equation of the first order in time derivative:
\begin{equation}
D\Pi  {+}
\Pi   n T \zeta \left[D\left(\frac{1}{n T \zeta} \right){-}9 \gamma_0 \right]  = 3 n T \zeta
D\left(\frac{\Theta}{n T} \right) \,,
\label{022}
\end{equation}
The procedure similar to the one used for the case $h_0 \neq 0$ gives the nonlinear equation of the second order for the energy density scalar
$$
x^2 W^{\prime \prime} {+}
xW^{\prime}\left[x \frac{W^{\prime}}{2W} {+} \frac{17{+}9\gamma}{2} \right] {+}
9(1{+}\gamma) W  =
$$
\begin{equation}
= \frac{3\zeta_0}{T} \left(x \frac{W^{\prime}}{2W} {+} 3 {-} x\frac{T^{\prime}}{T}   \right) {+}  \frac{9\zeta_0 \gamma_0 n_0}{\kappa x^3} \left[x\frac{W^{\prime}}{W} {+} 3(1{+}\gamma) \right].
\label{key01}
\end{equation}
With key equations (\ref{key1}) or (\ref{key01}) we are ready to analyze some exact solutions for the nonlocal model.

\section{Exact solutions to the key equations}

\subsection{Solution of the de Sitter type}

The set of master equations admits the solution with constant Hubble function $H=H_0$ in both cases $\gamma \neq -1$ and $\gamma=-1$. We consider these two cases separately.

\subsubsection{The sign of the parameter $\gamma$ is arbitrary, but $\gamma \neq -1$}

If we consider $W(x){=}W_{*}{=}const$, we obtain that $H{=}H_0 {=} \sqrt{\frac{\kappa W_{*}}{3}}$, and $a(t)= a(t_0)e^{H_0(t-t_0)}$, i.e., we deal with the solution of the de Sitter type, and the quantity $\kappa W_{*}$ plays the role of effective cosmological constant. If to put $W(x)=W_{*}$ to the key equation (\ref{key1}), we conclude that the temperature $T(x)$ has to satisfy the equation
\begin{equation}
x \left(\frac{1}{T}\right)^{\prime} {+}   \frac{3}{T} = 3(1{+}\gamma)  \left[ W_{*} \left(\frac{1}{\zeta_0} {+} \frac{9}{2} h_0  \right) {-}  \frac{3\gamma_0 n_0}{\kappa x^3} \right].
\label{Sitter1}
\end{equation}
The solution to this equation is
$$
\frac{1}{T(x)}= \frac{1}{x^3 T(1)} +  (1+\gamma) W_{*} \left(\frac{1}{\zeta_0} {+} \frac{9}{2} h_0  \right)\left(1-\frac{1}{x^3} \right) -
$$
\begin{equation}
-9 (1+\gamma) \frac{\gamma_0 n_0}{\kappa x^3}\log{x}  \,.
\label{Sitter2}
\end{equation}
Asymptotically, when $x \to \infty$, the inverse temperature tends to the constant
\begin{equation}
\frac{1}{T(\infty)}=  (1+\gamma) W_{*} \left(\frac{1}{\zeta_0} {+} \frac{9}{2} h_0  \right)
\,.
\label{asymSitter2}
\end{equation}
Mention should be made that the asymptotic value of the temperature does not depend on the nonlocal parameter $\gamma_0$.
Using the term $\frac{1}{T(\infty)}$ we can rewrite the expression (\ref{Sitter2}) as follows:
\begin{equation}
\frac{T(1)}{T(x)}= \frac{1}{x^3} {+}  \frac{T(1)}{T(\infty)}\left(1{-}\frac{1}{x^3} \right) {-} 9  (1{+}\gamma) T(1) \frac{\gamma_0 n_0}{\kappa x^3}\log{x}  \,.
\label{Sitter29}
\end{equation}
Taking into account that according to (\ref{T6}) we have for this case that
\begin{equation}
\frac{1}{T} = x^{3\gamma} {\cal H}(x^{-3\gamma})\,,
\label{T67}
\end{equation}
where ${\cal H}(z)$ is arbitrary function of its argument, one can represent the function ${\cal H}(z)$ as follows:
$$
{\cal H}(z) =  \frac{z}{T(\infty)} + \left[\frac{1}{T(1)}- \frac{1}{T(\infty)} \right] z ^{\frac{1+\gamma}{\gamma}} +
$$
\begin{equation}
+3\frac{(1+\gamma) \gamma_0 n_0}{\gamma \kappa} z^{\frac{1+\gamma}{\gamma }} \log{z}\,,
\label{Sitter3}
\end{equation}
In other words, when the temperature is presented by the admissible function (\ref{T67}), (\ref{Sitter3}), we obtain the exact solution of the nonlocally extended Israel-Stewart model, which can be characterized as the de Sitter type solution.

\subsubsection{Guiding parameters of the model}

Keeping in mind the solution (\ref{Sitter29}) we can extract three dimensionless parameters, which predetermine effectively the behavior of the temperature. The first one is the ratio $\frac{T(1)}{T(\infty)}$, which is formed using $W_*$, $\gamma$, $\zeta_0$, $h_0$ and $T(1)$; the second dimensionless parameter is  $T(1) \frac{\gamma_0 n_0}{\kappa}$; the third one is $(1{+}\gamma)$. For certainty, we assume that $T(1)>0$, $n_0>0$ and $h_0 > - \frac{2}{9\zeta_0}$, and, of course, $W_*>0$. Then the sign of the first parameter $\frac{T(1)}{T(\infty)}$ is predetermined by the sign of $(1{+}\gamma)$, and the sign of the second parameter coincides with the sign of the parameter $\gamma_0$. In this sense, we can indicate two parameters $(1{+}\gamma)$ and $\gamma_0$ as the guiding parameters of the model. We analyze the temperature law (\ref{Sitter29}) assuming sequentially that $1{+}\gamma>0$, $1{+}\gamma<0$ and $1{+}\gamma=0$. In all these cases we analyze the role of the parameter $\gamma_0$, which appears in the entropy flux four-vector (\ref{17}) together with the nonlocal term.

\subsubsection{Behavior of the inverse temperature at $\gamma>-1$ }

When $\gamma > -1$, according to (\ref{asymSitter2}) the asymptotic temperature $T(\infty)$ is positive; we assume that $T(1)>T(\infty)>0$. The derivative of the inverse temperature takes zero value, when $x=x_{(\rm ext)}$, satisfying the equality
\begin{equation}
3 \log{x_{(\rm ext)}} = 1 - \frac{\kappa}{3\gamma_0 n_0 (1+\gamma)}\left[\frac{1}{T(\infty)} - \frac{1}{T(1)}\right] \,.
\label{extrema1}
\end{equation}
Since the real value $x_{(\rm ext)}$ has to be more than one, $x_{(\rm ext)}>1$, it corresponds to the extremum of the function $\frac{T(1)}{T(x)}$, if the right-hand side of (\ref{extrema1}) is positive.
It is possible, either when $\gamma_0$ is negative (the extremum is the maximum), or when
\begin{equation}
 \gamma_0 \geq \frac{\kappa}{3 n_0 (1+\gamma)}\left[\frac{1}{T(\infty)} - \frac{1}{T(1)}\right] > 0 \,,
\label{extrema2}
\end{equation}
(the extremum is the minimum). When
\begin{equation}
 0 \leq \gamma_0 < \frac{\kappa}{3 n_0 (1+\gamma)}\left[\frac{1}{T(\infty)} - \frac{1}{T(1)}\right] \,,
\label{extrema2}
\end{equation}
the function $\frac{1}{T(x)}$ has no extrema.

Finally, when $T(1)=T(\infty)$, the extremal value of the variable $x$ is equal to $x_{(\rm ext)}=e^{\frac13}$; this value of the reduced scale function corresponds to the minimum, when $\gamma_0$ is positive, and to the maximum, when $\gamma_0$ is negative.

\subsubsection{Behavior of the inverse temperature at $\gamma<-1$ }

The authors of the paper \cite{B2} advocated the idea that the case $\gamma<-1$ is associated with the negative temperature, and that such interpretation of the dark fluid thermodynamics is admissible. If we would follow this idea and consider $T(1)<0$, $T(\infty)<0$, $T(x)<0$, $1+\gamma<0$ we could repeat the results obtained below using the replacements of these quantities by their moduli. We do not intend to analyze these results, but will focus the attention on the case $T(1)>0$, $T(\infty)<0$. Clearly, if the function $T(x)$ is continuous, a value $x_*$ should exist, for which the temperature takes zero value, $T(x_*)=0$. However, the formula (\ref{Sitter2}) does not admit such solution.

\subsubsection{The case $\gamma=-1$}

For this special case the solution for the temperature is $T(x) = T(1) x^3$. As it was emphasized in \cite{B2} the physically adequate solution has to be trivial $T(x)=T(1)=0$.

\begin{figure}[h!]
	\centering
	\includegraphics[width=\columnwidth]{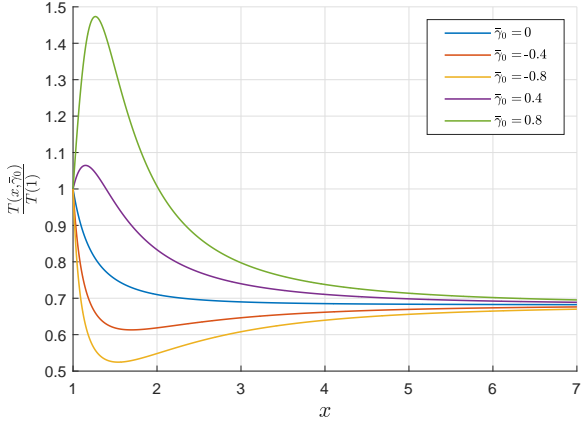}
	\caption{Illustration of the behavior of the reduced temperature $\frac{T(x,\bar{\gamma}_0)}{T(1)}$ for the de Sitter-type solution (\ref{Sitter29}) for the case $\gamma = -1/3 > - 1$. Here $x=\frac{a(t)}{a(t_0)}$ is the dimensionless scale factor, and $\bar{\gamma}_0 = \frac{\gamma_0n_0 T(1)}{\kappa}$ is the dimensionless parameter of non-locality. All the curves start with the value equal to one, and finish with the unified asymptotic value $\frac{T(\infty)}{T(1)}$. The basic curve related to $\bar{\gamma}_0=0$ is monotonic; when $\bar{\gamma}_0 \neq 0$, we find the extrema on the plots, which depend on the sign and value of $\bar{\gamma}_0$ (they are indicated in the upper right corner of the panel).}
	\label{fig1}
\end{figure}

\subsection{Power-law type solutions}

\subsubsection{The key equation for the energy density scalar}

The key equation is nonlinear in the function $W$ and in
its first derivative $W^{\prime}(x)$. In order to find its exact particular solutions we assume the energy density scalar to have the power - law form $W=W_{*} x^{\lambda}$.  Now the key equation (\ref{key1}) converts into
$$
\frac32 x^{\lambda} h_0 W_{*} \left\{\lambda^3 + \lambda^2\left[\frac12(13 + 6\gamma) + \frac{1}{3\zeta_0 h_0} \right] +  \right.
$$
$$
\left. + \lambda\left[\frac12(27 {+} 21\gamma) {+} \frac{5{+}3\gamma}{3\zeta_0 h_0} \right]   {+} 2(1+\gamma) \left(\frac92 {+} \frac{1}{\zeta_0 h_0} \right)\right\} =
$$
\begin{equation}
= x \left(\frac{1}{T}\right)^{\prime} + \left(3+ \frac{\lambda}{2}\right) \left(\frac{1}{T}\right) {+}  \frac{3\gamma_0 n_0}{\kappa x^3}  \left[\lambda {+} 3(1{+}\gamma) \right].
\label{key78}
\end{equation}
When $h_0 =0$, this equation can be rewritten as
$$
\frac{1}{2\zeta_0} x^{\lambda} W_{*} \left[\lambda^2 +  \lambda (5+3\gamma)   + 6(1+\gamma) \right] =
$$
\begin{equation}
= x \left(\frac{1}{T}\right)^{\prime} + \left(3+ \frac{\lambda}{2}\right) \left(\frac{1}{T}\right) {+}  \frac{3\gamma_0 n_0}{\kappa x^3}  \left[\lambda {+} 3(1{+}\gamma) \right]\,.
\label{key789}
\end{equation}
The non-equilibrium pressure $\Pi$ can be written now as
 \begin{equation}
\Pi(x)= P(x) \frac{1}{\gamma} \left[\lambda + 3(1+\gamma) \right]\,,
\label{pi11}
\end{equation}
i.e., the ratio $\frac{\Pi(x)}{P(x)}$ is constant and is predetermined by the parameters $\lambda$ and $\gamma$.

\subsubsection{Evolution of the effective temperature}

Our ansatz is that the effective temperature satisfies the equation
\begin{equation}
x \left(\frac{1}{T}\right)^{\prime} {+} \left(3+ \frac{\lambda}{2}\right) \left(\frac{1}{T}\right) {+}  \frac{3\gamma_0 n_0}{\kappa x^3}  \left[\lambda {+} 3(1{+}\gamma) \right] = 0\,.
\label{temp1}
\end{equation}
This ansatz assumes that the guiding parameter $\gamma_0$ enters the equation for the temperature, but does not appear in the left-hand sides of the of the equations (\ref{key78}) and (\ref{key789}). The solution to the equation (\ref{temp1}) is
\begin{equation}
\frac{1}{T(x)}= x^{-3}\left\{ \frac{x^{-\frac{\lambda}{2}}}{T(1)} {+} \frac{6\gamma_0 n_0}{\kappa \lambda}\left[\lambda {+}3(1{+}\gamma) \right]\left[x^{-\frac{\lambda}{2}} {-} 1 \right]\right\}.
\label{Sitter342}
\end{equation}
As for the admissible function $\tilde{\cal H}(z)$ associated with this solution, we have to use the formula
\begin{equation}
\frac{1}{T(x)} = x^{3\gamma} \tilde{\cal H}\left(x^{\frac{\gamma}{1+\gamma}\left[\lambda+3(1+\gamma) \right]} \right) \,,
\label{gamma99}
\end{equation}
where $\tilde{\cal H}(z)$ is again an arbitrary function of its argument. We can find now that
$$
\tilde{\cal H}(z) =
\left\{\frac{1}{T(1)} {+} \frac{6\gamma_0 n_0}{\kappa \lambda} \left[\lambda {+}3(1{+}\gamma) \right]\right\} z^{-\frac{1{+}\gamma}{2\gamma}\left[\frac{\lambda{+}6(1{+}\gamma)}{\lambda{+}3(1{+}\gamma)}\right] } {-}
$$
\begin{equation}
-\frac{6\gamma_0 n_0}{\kappa \lambda} \left[\lambda+3(1+\gamma) \right] z^{-\frac{3(1+\gamma)^2}{\gamma \left[\lambda+3(1+\gamma)\right]}} \,.
\label{gamma919}
\end{equation}
Asymptotically, at $x \to \infty$ we obtain that $T \to 0$, if $\lambda< -6$. When $\lambda=-6$, the law for the inverse temperature is
\begin{equation}
\frac{1}{T(x)}= \frac{1}{T(1)} {+} \frac{3\gamma_0 n_0}{\kappa}(1{-}\gamma) \left(1 - x^{-3} \right) \,,
\label{SitterTT}
\end{equation}
and thus the asymptotic value
\begin{equation}
\frac{1}{T(\infty)}= \frac{1}{T(1)} {+} \frac{3\gamma_0 n_0}{\kappa}(1{-}\gamma) \,,
\label{SitterTT6}
\end{equation}
is a constant depending on the guiding parameters $\gamma$ and $\gamma_0$. In particular, when $\gamma_0(1-\gamma)>0$, we obtain that $T(\infty)<T(1)$. When $\gamma=1$, i.e. the equilibrium state is characterized by the strict equation of state, $T(\infty)=T(1)=T(x)$, i.e., the temperature is constant.

\subsubsection{Geometric properties of the model}

The Hubble function  $H(x)$ found in terms of the reduced scale factor $x$
\begin{equation}
H(x) = \sqrt{\frac{\kappa W_*}{3}} x^{\frac{\lambda}{2}} \,,
\label{H1}
\end{equation}
allows us to recover the dependence $a(t)$ by the standard formula
\begin{equation}
t-t_0 = \int_1^{\frac{a(t)}{a(t_0)}} \frac{dx}{xH(x)}\,.
\label{H2}
\end{equation}
We obtain immediately that
\begin{equation}
\frac{a(t)}{a(t_0)} = \left[1- \frac{\lambda}{2}(t-t_0)\sqrt{\frac{\kappa W_*}{3}} \right]^{-\frac{2}{\lambda}} \,.
\label{H3}
\end{equation}
In terms of the cosmological time the Hubble function is
\begin{equation}
H(t) = \sqrt{\frac{\kappa W_*}{3}} \left[1- \frac{\lambda}{2}(t-t_0)\sqrt{\frac{\kappa W_*}{3}} \right]^{-1} \,.
\label{H4}
\end{equation}
The acceleration parameter has now the form
\begin{equation}
-q \equiv \frac{\ddot{a}}{a H^2} = 1+ \frac{\lambda}{2} \,.
\label{H5}
\end{equation}
Universe expands with acceleration, if $\lambda>-2$.

\subsubsection{Characteristic equation and its solutions}

When the temperature satisfies the equation (\ref{temp1}), the parameter $\lambda$ has to satisfy the cubic characteristic equation
$$
\lambda^3 + \lambda^2\left[\frac12(13 + 6\gamma) + \frac{1}{3\zeta_0 h_0} \right] +
$$
\begin{equation}
 {+} \lambda\left[\frac12(27 {+} 21\gamma) {+} \frac{5{+}3\gamma}{3\zeta_0 h_0} \right]   {+} 2(1{+}\gamma) \left(\frac92 {+} \frac{1}{\zeta_0 h_0} \right) = 0 \,.
\label{key784}
\end{equation}
Depending on values of two parameters $\gamma$ and $\xi \equiv \frac{1}{3\zeta_0 h_0}$ this equation can have one real solution plus two complex conjugated, or three real solutions $\lambda_1$, $\lambda_2$, $\lambda_3$. These real roots can be different, say, $\lambda_1>\lambda_2>\lambda_3$;  two of them can coincide, $\lambda_1{=}\lambda_2 \neq\lambda_3$, or three roots can coincide $\lambda_1{=}\lambda_2{=} \lambda_3$. These solutions can be standardly presented by the Cardano formulas and describe various regimes of the cosmological expansion.

As an illustration, we consider only one special case $\gamma=-1$, for which the characteristic equation
\begin{equation}
\lambda^3 + \lambda^2\left(\frac72 + \xi \right) +  \lambda\left(3+2\xi \right)   = 0
\label{char6}
\end{equation}
has the following roots:
\begin{equation}
\lambda_1 =0 \,,\quad \lambda_2 = -2\,, \quad \lambda_3 = - \frac12 (3+2\xi) \,.
\label{zero1}
\end{equation}
Clearly, the root  $\lambda_1 {=}0$ converts the power-law solution to the de Sitter one.
Finally, when $h_0=0$, the characteristic equation (\ref{key784}) converts into the quadratic equation
\begin{equation}
 \lambda^2 + \lambda (5{+}3\gamma)   {+} 6(1{+}\gamma) = 0 \,,
\label{key787}
\end{equation}
the solutions to which are $\lambda_1=-2$, $\lambda_2 = -3(1+\gamma)$.

\subsubsection{Behavior of the temperature: An example}

Let us consider one analytic example of the temperature behavior for the case $\gamma=-1$ and $\lambda = -8$ (it can be obtained from the root $\lambda_3$ in (\ref{zero1}), if $\xi = \frac{13}{2}$).
The corresponding solution for the inverse temperature can be written as
\begin{equation}
\frac{T(1)}{T(x)} =  \left[1+ \frac{6\gamma_0 n_0 T(1)}{\kappa} \right] x -   \frac{6\gamma_0 n_0 T(1)}{\kappa} x^{-3}  \,.
\label{example1}
\end{equation}
When the guiding parameter $\gamma_0$ is positive, this function grows monotonically from one to infinity, so that the corresponding temperature tends asymptotically to zero as $T(x) \propto \frac{1}{x}$. This asymptotic law rewritten as $T a = const$ is similar to the behavior of the temperature of the cosmic microwave background radiation.

When $\gamma_0$ is negative, and it satisfies the inequality
\begin{equation}
\frac14 < \frac{6 |\gamma_0| n_0 T(1)}{\kappa} <1  \,,
\label{example2}
\end{equation}
the plot of the function $\frac{T(1)}{T(x)}$ has the minimum at $x{=}x_{(\rm m)}$, where
\begin{equation}
x_{(\rm m)} = \left\{\frac13 \left[\frac{\kappa}{6 |\gamma_0| n_0 T(1)} -1 \right]  \right\}^{-\frac14} \,.
\label{example2}
\end{equation}
This means that starting from $T(1)>0$ the temperature itself, $T(x)$ grows, reaches the maximum at $x{=}x_{(\rm m)}$ and then tends to zero asymptotically.

\begin{figure}[h!]
	\centering
	\includegraphics[width=\columnwidth]{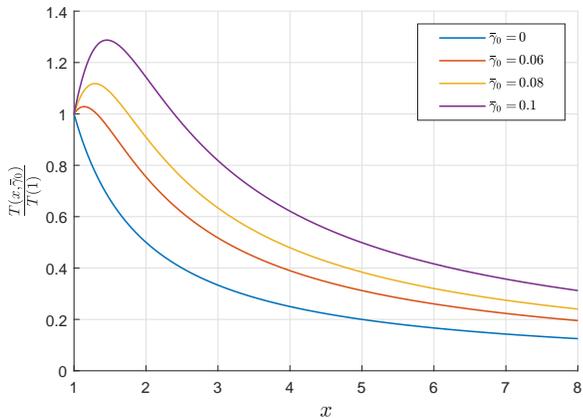}
	\caption{Illustration of the behavior of the reduced temperature $\frac{T(x,\bar{\gamma}_0)}{T(1)}$ for the power-law solution (\ref{example1}). The basic curve related to $\bar{\gamma}_0=0$ is monotonic; when $\bar{\gamma}_0 \neq 0$, there are extrema depending on the sign and values of $\bar{\gamma}_0$.}
	\label{fig1}
\end{figure}

\section{Conclusions}

We presented the nonlocal extension of the Israel-Stewart causal thermodynamics of the cosmic fluid, which inherits the isotropy and homogeneity of the expanding Universe. We would like to formulate briefly the following results of the analysis of the model as a whole, and of the exact solutions, which we obtained for two cases: the de Sitter type  and the power-law solutions.

\noindent
1. The established nonlocal extension of the Israel-Stewart causal thermodynamics is the relativistic analog of the Burgers model known in classical viscoelasticity (please, compare (\ref{Burgers1}) with (\ref{cc1})).

\noindent
2. For the linear barotropic equation of state of the cosmic fluid we reconstructed the properties of the effective temperature (\ref{T6}).

\noindent
3. The de Sitter type solution is provided by the specific choice of the effective temperature (\ref{Sitter2}), which  tends asymptotically to the constant; the effective cosmological constant is presented by the value $\kappa W_*$, where $W_*$ is a stationary value of the energy density of the cosmic fluid; the de Sitter type solution coincides with power-low solution, when the value $\lambda =0$ happens to be the root of the characteristic equation (see, e.g., (\ref{char6})).

\noindent
4. The parameter $\lambda$, which predetermines the properties of the power-law solution of the model, satisfies the cubic characteristic equation (\ref{key784}), and thus can describe at most three stages in the Universe history, which correspond to three real roots of the characteristic equation; the value of this parameter depends on two constants: the effective bulk viscosity $\zeta_0$ and the effective relaxation parameter $h_0$.

\noindent
5. For the case $\lambda<0$ the scale factor of the expanding Universe is monotonic and is presented by the power-law function (\ref{H3}).

\noindent
6. The effective temperature vanishes asymptotically, when the parameter $\lambda$ satisfies the inequality $\lambda<-6$ (see, e.g., (\ref{Sitter342})).

\noindent
7. The sign and the value of the guiding parameter $\gamma_0$, which appears in front of the nonlocal term in the entropy flux four-vector (\ref{17}), predetermines the presence or absence of extrema in the plots of the fluid temperature (see Figs 1,2).

\acknowledgments{The work was supported by the Russian Science Foundation (Grant No 21-12-00130), and partially by the Kazan Federal University Strategic Academic Leadership Program.}

\end{document}